\documentclass[12pt]{article}

\usepackage{graphicx}
\usepackage{a4wide}
\usepackage{mathrsfs}
\usepackage{amsmath}
\usepackage{amsfonts}

\begin{document}

\title{Kaluza-Klein Contamination in Fermi Accelerated Environments}
\author{Cong-Xin Qiu\\
\small{Department of Astronomy, Nanjing University, 22 Hankou Road}\\
\small{Nanjing, Jiangsu 210093, P. R. China}\\
\small{http://oxo.lamost.org/}\\
\small{congxin.qiu@gmail.com}
}
\date{~}

\maketitle

\begin{abstract}
Astrophysical constraints of new physics are often limited to weakly
interacting light particles, such as axions, the Kaluza-Klein (KK)
gravitons from the ADD model, sterile neutrinos and unparticles. We
discuss the possibility for an astrophysical scenario to
(dis)confirm new physics for heavy particles beyond $\mathrm{TeV}$
energy scale. In our scenario, the KK protons (the KK excited
quarks/gluons within protons) within the framework of universal
extra dimensions (UEDs), are produced by high energy $p + p$
collisions in Fermi accelerated environments, with protonic
isotropic spectrum $d N / d E \propto E^{-2}$ up to at least
$10^{18}\,\mathrm{eV}$. Thus, because they are also electrically
charged, they should be re-accelerated by mechanism similar to
normal protons. The KK states (no matter whether they have already
decayed to the lightest KK particle or not) should contaminate
$10^{-5}$ to $10^{-2}$ of cosmic-ray events for some fixed energy
$E$ (within some suitable assumptions). Hence, if we have techniques
to identify them from air shower data, we can constrain UEDs
scenario. Our method is an ``existence proof'' that we can constrain
new physics beyond $\mathrm{TeV}$ scale or much higher by classical
astrophysical scenarios, which can also be generalized to
supersymmetric models, the bulk Standard Model fields within the RS
model, and the endlessly emerging new models. Moreover, it can
exploit domains which have no possibility to be studied in
terrestrial experiments.

\vspace{0.5cm} \noindent
\end{abstract}

\section{Introduction~\label{(sec)_introduction}}

Brane-world scenarios, such as the Arkani-Hamed-Dimopoulos-Dvali
(ADD)
model~\cite{ArkaniHamed:1998rs,Antoniadis:1998ig,ArkaniHamed:1998nn}
and Randall-Sundrum (RS) model~\cite{Randall:1999ee,Randall:1999vf},
give an alternative framework for solving the hierarchy problem.
Within the ADD model, the Standard Model (SM) fields are confined to
a $3$-brane ($(3+1)$-dimensional spacetime) while gravitons
propagate freely in a torus compactified bulk space (large extra
dimensions), and the gravitational coupling constant $G =
1/M_{\mathrm{pl}}^2$ observed in our $(3+1)$-dimensional world is an
effective one. So it is natural to understand why $G$ is so small.
The RS model solves the same problem by a slice of $AdS_5$
spacetime. Some superstring-inspired descriptions make these models
more attractive.

Some lineage scenarios of the ADD model also allow the SM particles
propagating, to some extent, in the bulk spaces. Beside gravitons,
such kind of SM particles can also be Kaluza-Klein (KK) excited in
the extra dimensions, thus give abundant physical phenomena. A
natural extension of the ADD model is let the brane with a finite
thickness and complex substructures~\cite{DeRujula:2000he}. An
example is given in Ref.~\cite{ArkaniHamed:1999dc}. In this
scenario, quarks and leptons are confined to different branes, while
the Higgs and SM gauge fields are sandwiched in, hence the
possibility for proton decay can be exponentially suppressed. The KK
excited gauge particles (which may be called branons) can be either
baryophobic or leptophobic, because they feel nontrivial on the
brane substructures. Another natural extension is asymmetrical
compactification, which has two (as minimum, maybe more) separate
compactification scales. The ``very large'' extra dimensions let
only graviton propagate, just as what the ADD model says, but the
$\mathrm{TeV}^{-1}$ scale ``large'' extra dimensions, may have the
SM fields extending~\cite{Lykken:1999ms}. Of course we can on the
other hand keep all compact dimensions at $\mathrm{TeV}^{-1}$ scale
rather than make the compactification asymmetric. However, the
advantage of solving the hierarchy problem in the ADD model is
bereft.

In the case of that the SM fields also extend to some extra
dimensions, to obtain chiral fermions in the 4-dimensional effective
theory, we have only two ways to go: (i) to confine fermions to
branes only~\cite{Dicus:2000hm}, or (ii) to impose bulk fermions
orbifold boundary conditions~\cite{Georgi:2000wb}. Universal extra
dimensions (UEDs)~\cite{Appelquist:2000nn} scenario is an example of
the second approach. In UEDs scenario, all SM fields can propagate
in these ``universal'' extra dimensions, and conservation of
momentum in the universal dimensions turns to conservation of the KK
number in our $(3+1)$-dimensional world. For two or more universal
extra dimensions, the na\"{i}ve KK mode sums diverge when the KK
tower number $N_\mathrm{KK} \rightarrow \infty$. So, let us consider
only one universal extra dimension as minimal universal extra
dimensions (MUEDs) in this context. In this case, $S^1 /
\mathbb{Z}_2$ orbifold compactification is assumed, and the KK mass
eigenvalues have a simple form $M_n^\mathrm{KK} = n / R$, where $R$
is the compactification scale. For the reason that only loop
diagrams can contribute electroweak observables by the KK number
conservation, the experimental bound for UEDs is only
$M_1^\mathrm{KK} = 1/R \geq 300\,\mathrm{GeV}$. In the tree level,
the mass spectrum of the KK excited SM particles has the form
$M_{\mathrm{SM},\,n} = \sqrt{(M_n^\mathrm{KK})^2 +
M_\mathrm{SM}^2}$, where $M_\mathrm{SM}$ is the zero-mode on-shell
mass of the corresponding SM particles. Hence, their masses are
level-by-level highly degenerated when $M_\mathrm{SM} \ll
M_n^\mathrm{KK}$. However, when radiative corrections are
concerned~\cite{Georgi:2000ks,Cheng:2002iz}, the mass degeneration
is broken, to some extend, as $M_{\mathrm{g},\,n}
> M_{\mathrm{Q},\,n} > M_{\mathrm{q},\,n} > M_{\mathrm{W^\pm},\,n}
\sim M_{\mathrm{Z^0},\,n} > M_{\mathrm{L},\,n} > M_{\mathrm{l},\,n}
> M_{\mathrm{\gamma},\,n} \sim M_n$, where $g$ denotes gluon, $Q$ ($L$) denotes
weak-doublet quark (lepton), $q$ ($l$) denotes weak-singlet quark
(lepton), and $\gamma$ denotes photon. The KK number conservation
breaks down to a KK parity that the even and odd KK numbers cannot
transform to each other. The correction scale depends on some
unknown parameters; however, some reasonable choice of parameters
shows that the largest correction $\Delta M_{\mathrm{g},\,n}$ may be
as large as $10\%$~\cite{Cheng:2002iz}. Hence, heavier KK excited
states should cascade decay (by the KK conserving or even violating
interactions) to the lightest KK particle (LKP) $\gamma_1$ which is
stable~\cite{Cheng:2002ab}, by emitting soft SM particles. When
considering the possibilities of experimental discovery, one always
assumes that the lifetimes of heavier KK excited states are
sufficiently short, thus the states can decay within the collider;
however, it is not supposed to do so. The total width cannot in fact
be calculated by the failure of reconstructing the Breit-Wigner
resonance~\cite{Datta:2005zs}. Notice that the decay rate of an
unstable particle $d\Gamma \propto 1/m_\mathcal{A}$ in the phase
space formula, the lifetime $\tau \propto m_\mathcal{A}$ where
$m_\mathcal{A}$ is the mass of the decay particle. If the $\Delta
M$s are smaller for a set of parameters different from
in~\cite{Cheng:2002iz}, or soft SM cascade processes are suppressed
by other reasons, the lifetimes of heavier KK excited states should
be even longer. Specific calculations for whether the
not-very-short-lived KK excited states can affect other more mature
scientific scenarios, such as disturb predictions of Big-bang
nucleosynthesis, or distort the cosmic microwave background, are
need; however, they are beyond our scope of our paper. Some
na\"{i}ve considerations show that all of them are not very crucial,
because we do not really need longer-lived KK excited states
(although the long-lived ones are also possibilities we shall
consider in the identification section in
\S\ref{(sec)_air_shower_identification}), but some
not-very-short-lived KK excited states to suffer the time scale of
Fermi acceleration (which is maybe $\sim \mathrm{s}$ or much
shorter), which is much shorter than the time scale of the scenarios
we mentioned above. So we assume that the lifetimes of heavier KK
excited states are long enough to suffer the astrophysical scenario
we draw in this paper.

Astrophysical constraints of new physics are often limited to weakly
interacting light particles, such as
axions~\cite{Dicus:1979ch,Fukugita:1982gn,Iwamoto:1984ir,Dearborn:1985gp,Frieman:1987ui,Raffelt:1987yu,Raffelt:1987yt,Turner:1987by,Burrows:1988ah,Haxton:1991pu,Raffelt:1994ry,Keil:1996ju},
the KK gravitons from the ADD
model~\cite{ArkaniHamed:1998nn,Barger:1999jf,Cullen:1999hc,Cassisi:2000hy,Hanhart:2000er,Biesiada:2001iy,Hannestad:2001xi,Hannestad:2003yd},
sterile
neutrinos~\cite{Kusenko:1997sp,Kusenko:1998bk,Hidaka:2006sg,Fryer:2005sz}
and unparticles~\cite{Hannestad:2007ys}. We want to construct an
astrophysical scenario to (dis)confirm new physics for heavy
particles beyond $\mathrm{TeV}$ energy scale. Notice that in Fermi
accelerated environments, protons in a power law spectrum up to at
least $10^{18}\,\mathrm{eV}$ should be produced (even if we have
already derived an overall Lorentz factor $\Gamma \sim 300$), thus
$p + p$ collisions up to a tremendously large energy should happen
there, which we cannot even imagine in terrestrial experiments.
However, we have to brain storm to know their happenings. In this
paper, we construct a scenario which may (dis)confirm UEDs by high
energy observation of cosmic-rays. This scenario may or may not have
opportunities to give stronger bounds than colliders, because of the
large uncertainties in our estimations, and the technical details of
lots of synergic scientific domains (which we cannot discuss at
length in this paper). However, it is at least an ``existence
proof'' for this kind of methodology. It can also explore domains
which have no possibility to be studied in terrestrial experiments.
Some similar scenario in Ref.~\cite{Anchordoqui:2004bd}, also
suggested the production of some kind of strongly interacting
massive particles in $p + p$ collisions in astrophysical
environments; however, our scenario have a lot of advantages than
theirs. The advantages rise mainly because (i) one of the protons in
their scenario stays at rest, but both of the protons in our
scenario are Fermi accelerated, and (ii) our KK excited states
suffer an additional accelerated process. Detail comparisons are
given in \S\ref{(sec)_discussion_outlook}.

In our scenario, the KK protons (with either KK excited quarks or
gluons in it) are produced by $p + p$ collisions in Fermi
accelerated environments. Both the original Fermi mechanism or
diffusive shock accelerating model have an isotropic spectrum $d N /
d E \propto E^{-2}$ up to at least $10^{18}\,\mathrm{eV}$, hence
they are okay for our purpose. The KK protons should also be
accelerated just as normal protons by the same mechanism; however,
they should have different properties than normal ones. Beside being
discovered one by one from air shower data directly, they may be
accelerated to energies normal protons cannot be accelerated to, or
they (or their decayed final state) may contaminate significant
amount of ultra-high-energy cosmic-ray (UHECR) events because of an
overall energy shift, both of which may make them a discovery. In
\S\ref{(sec)_Producing}, we calculate the cross section and
production rate of the KK protons. We show that the production rate
may be large enough to make meaningful scientific constraints. In
\S\ref{(sec)_accelerating_propagating}, we discuss the accelerated
property of them in Fermi accelerated environments, making a
comparison with normal protons. We notice that the KK states should
contaminate $10^{-5}$ (for special sources) to $10^{-2}$ (for
diffuse flux) of cosmic-ray events for some fixed energy $E$ (if
assuming the optical depth $\tau_\mathrm{pp} = 1$), which are not
too small a sample to be discovered by air shower detection. We also
discuss the propagating properties of them related to soft photon
interactions. In \S\ref{(sec)_air_shower_identification}, we
consider the probabilities to identify them (or their decayed final
state) from other cosmic-ray particles from air shower data. If it
can be done so, our method can have larger possibilities to give
smaller parameter space for UEDs than other methods. In
\S\ref{(sec)_neutrino_detectors}, we calculate the possible
constraints of the KK cosmic-ray flux from neutrino detectors;
however, the constraints are very loose for current scientific
equipments to affect our former estimations. We discuss our results
and draw the possibilities to generalize our method to other new
physics models in \S\ref{(sec)_discussion_outlook}.

\section{Producing of the KK Protons\label{(sec)_Producing}}

The KK particles can be produced by high energy collisions, both in
terrestrial experiments and astrophysical environments. We may want
to consider a list of some not-very-exotic processes, such as
$\gamma + \gamma$, $e + e$, $p + p$, $p + n$, $e + \gamma$ and $p +
\gamma$. Collisions including electrons are very tempting; however,
electrons always have limited energy because of synchrotron
radiation. Ultra-high-energy (UHE) photons with energy
$10^{12-14}\,\mathrm{eV}$ or larger may be produced in astrophysical
scenarios~\cite{Gupta:2007yb}; however, we leave behind this
possibility elsewhere because of page limitations. Collisions
including neutrons are very interesting; however, uncharged neutrons
are always less energetic than protons, thus they make a less
center-of-mass energy $\sqrt{s}$ for $p + n$ interactions than $p +
p$ ones. In addition, we lack trustworthy parton distribution
functions (PDFs) $f(x, Q)$ for high $Q$ neutrons. Hence, we will
only discuss $p + p$ collisions in this paper.

\subsection{$p + p \rightarrow \mbox{(the KK states)}$ Cross Sections\label{(subsec)_p+p}}

The $p + p$ cross sections to produce the KK bosons and fermions in
the framework of UEDs have already been calculated
in~\cite{Rizzo:2001sd,Macesanu:2002db}. However, their motivation is
mainly on whether we have possibilities to confirm them on
synchrotrons (especially Tevatron I, II and the Large Hadron
Collider (LHC)), thus they focused their calculations on a
center-of-mass energy $\sqrt{s}$ of $1\,\mathrm{TeV}$,
$2\,\mathrm{TeV}$ and $14\,\mathrm{TeV}$ respectively and let
$M_1^\mathrm{KK}$ as a free parameter. As a result, their
calculations are not suitable for our purpose. So we redo the
calculations for a set of parameters $\sqrt{s}$. In this paper, we
always fix $M_1^\mathrm{KK} = 1/R = 350\,\mathrm{GeV}$ for
simplicity.

In our calculation, we assume the KK states are sufficiently stable
(see the discussions in \S\ref{(sec)_introduction}). We use the
amplitude-squared $\overline{\Sigma|}\mathcal{M}|^2$ calculated by
Macesanu et al.~\cite{Macesanu:2002db}. Their expression for total
cross section is
\begin{equation}
\begin{split}
    \sigma_\mathrm{tot}^\mathrm{KK} = &
        \frac{1}{4 \pi} \sum_j \sum_n
        \int_{\rho_n}^1 d x_A \int_{\rho_n/x_A}^1 d x_B \\
        &\times f_1(x_A, Q) f_2(x_B, Q)
        \times \int_{0}^{\pi} \sin \theta d \theta
        \,\frac{\overline{\Sigma|}\mathcal{M}_j|^2}{S!}
        \frac{1}{s} \sqrt{1 - \frac{4 M_n^2}{s}}
        \mbox{,}
\end{split}
\end{equation}
where $S$ is a statistical factor to memorize the number of
identical final states, $s$, $t(u) = - (s/2) (1 - M_n^2/s) (1 \mp
\cos\theta)$ are the Mandelstam variables, $\theta$ is the angle
between two incident particles in the source comoving frame. We
evaluate the PDFs $f(x, Q)$ by CTEQ6.6m (standard MSbar
scheme)~\cite{Pumplin:2002vw,Nadolsky:2008zw}. The calculation of
total cross section $\sigma_\mathrm{tot}^\mathrm{KK}$ is too
cumbersome although routinely, because there are too many
subprocesses and too many quark flavors. Notice that (when $Q =
350\,\mathrm{GeV}$) u-quarks dominate in the large $x$ region and
gluons dominate in the small $x$ region. So we neglect the
subprocess for initial states s, c, b and anti-quarks completely in
the context. We also neglect the contribution of d-quarks because
their PDFs are always smaller than u-quarks; however, this reduction
may not be suitable for arbitrary $Q$. Hence, we will only calculate
subprocess including u-quarks and gluons. In our calculation,
$M_{\mathrm{SM},\,n} = M_n^\mathrm{KK}$ is assumed for simplicity,
for the reason that both zero-mode on-shell mass and radiative
corrections cannot alter the masses of magnitude.

\begin{figure}[ht]
\begin{center}
\includegraphics[angle=0,width=8cm]{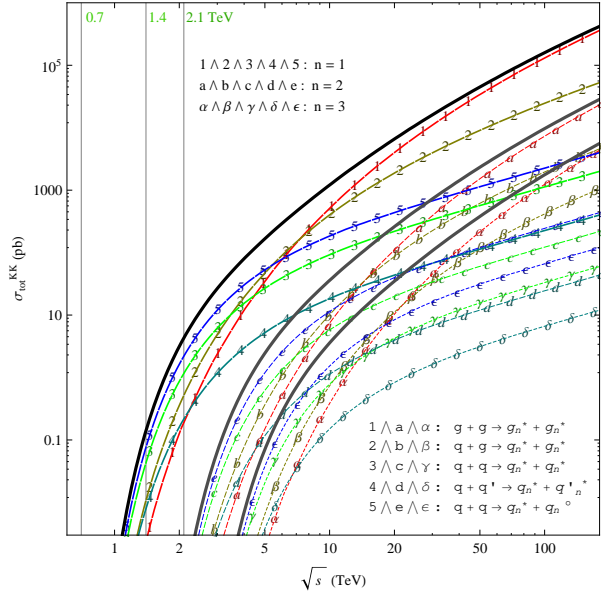}
\end{center}
\caption{Cross section for the case $n = 1, 2, 3$ and
subprocess $g + g \rightarrow g_n^\ast + g_n^\ast$, $u + g
\rightarrow u_n^\ast + g_n^\ast$, $u + u \rightarrow u_n^\ast +
u_n^\ast$, $u + u^{'} \rightarrow u_n^\ast + u_n^{'\,\ast}$ and $u +
u \rightarrow u_n^\ast + u_n^\circ$ (Eq.~(15), (17), (18), (22) and
(24) in Ref.~\cite{Macesanu:2002db} respectively), as marked in
the figure. The total cross sections for different $n$ are showed by
thick black lines. We have assumed $q = u$ because u-quarks dominate
the quark PDFs in a large range of energy scale $Q$. The other
process $g + g \rightarrow q_n^\ast + \bar{q}_n^\ast$ may be
important; however, we neglect it because of a lack of consistency
amplitude-squared.} \label{(fig)_cross_section}
\end{figure}

The cross section is calculated in Fig.~\ref{(fig)_cross_section}.
We integrate out the $\int d \theta$ to neglect the transverse
momentum distribution, which is irrelevant to astrophysical
scenarios. We use only the lowest order (LO) expression for quantum
chromodynamics (QCD) running coupling constant $\alpha_S(Q)$ in the
amplitude-squared, which is sufficient for high energy regions $Q
\gg M_{Z^0}$. We also assume $Q = M_n$~\footnote{However, this
assumption may need careful considerations when $\sqrt{s}$ is much
larger than $M_n$. We assume it because a refined calculation is far
beyond our scope in this paper.} for simplicity. We do not show the
uncertainties of the PDF sets in that figure; however, some
tentative calculations show that error bars can be neglectable for a
large range of $Q$, and it will never affect our rough astrophysical
estimations. As seen from the figure, the contribution of $n = 1$
state is always more important than higher excited states in the
calculable regions, thus we neglect the others. The subprocess $g +
g \rightarrow g_n^\ast + g_n^\ast$ dominate the large $\sqrt{s}$
region, and $q + q \rightarrow q_n^\ast + q_n^\circ$ dominate the
small $\sqrt{s}$ region. In magnitude, our results are consistent
with~\cite{Macesanu:2002db} in the regions not adjacent to $(2
M_n)_+$. However, for energy just above the threshold $\sqrt{s} = (2
M_n)_+$, our $\sigma$s are largely suppressed rather than mildly
decreased in theirs. In addition, it seems that their
$\sigma^\mathrm{KK}(g + g \rightarrow)$ increases more softly than
ours. These may only be caused by some numerical details, so we will
not discuss these issues deeply. Hence $p + p$ collisions can
produce the KK protons with the KK quarks or gluons in it (we will
call them by a joint name $p_\mathrm{KK}$). The reason why the
subprocess $g + g \rightarrow g_n^\ast + g_n^\ast$ is also suitable
for our purpose to produce the KK excited protons is that in the
high energy collisions gluon bremsstrahlung is unsuppressed, thus
the two gluons will produce two jet events to form two protons.
Because we have already assumed that the KK number conservation is
exact in UEDs scenario, a KK excited gluon cannot take off its KK
number and cannot leave the proton for color confinement. Thus both
the KK excited bosons and fermions can form the KK excited protons.

The calculated cross section result has a lot of cumbersome
difficulties, mainly in the high $\sqrt{s}$ regions (it has to
apply) up to at least $10^{18}\,\mathrm{eV}$. The difficulties are
listed as follows: (i) Because the PDFs are only applicable to
$\rho_n = 4 M_n^2 / s
> x_\mathrm{min} = 10^{-8}$ in CTEQ6.6 series, we can only calculate
cross sections below energy $\sqrt{s_\mathrm{max}} = 7 \times
10^{15}\,\mathrm{eV}$~\footnote{We cannot saturate the bound because
of our limited computation power. However, increasing the
calculation load considerably can have little improvement on the
result of $E_\mathrm{max}$, thus we do some qualitative analyses
instead in \S\ref{(subsec)_High_Energy_Behavior}.} for the lowest KK
state $M_1^\mathrm{KK} = 350\,\mathrm{GeV}$, which is insufficient
for astrophysical events that happen in Fermi accelerated
environments. (ii) Multi-KK processes (just like the multi-$\pi$
processes in soft hadron physics) have to be considered when
$\sqrt{s}$ is much larger than $M_n$. (iii) The Lagrangian we use to
calculate the cross sections may only be an effective one (thus
non-renormalizable in the framework of quantum field theory), hence
cannot be applied to $\sqrt{s}$ larger than some cutoff energy scale
(which may be not very larger than $M_1$). Problem (i) and (ii) are
the main topics of \S\ref{(subsec)_High_Energy_Behavior}. Problem
(iii) is much more troublesome, thus we can only solve it completely
after the emergence of the final theory (which should explain what
is the physical essence of normalization). We assume their
capabilities simply because (i) some smoothness considerations, and
(ii) we do not have other ways to go. Other authors used the same
strategies as ours, just like what in
Ref.~\cite{AlvarezMuniz:2001mk}. Moreover, because of the
re-acceleration of our KK particles, the energetic KK ones need not
to be produced by high $\sqrt{s}$ $p + p$ collisions; hence, the
high $\sqrt{s}$ behaviors of our cross sections may in fact not very
crucial. It should be that case at least for the na\"{i}ve
estimations we choose in this paper, because we can hardly
distinguish the difference between the KN and DL parametrizations in
Fig.~\ref{(fig)_KK_production_Fermi}; however, we do not know the
applicability of this supposition. We leave the more quantitative
discussions in the follow-up publications.

\subsection{High Energy Behavior\label{(subsec)_High_Energy_Behavior}}

As we have already mentioned before in \S\ref{(subsec)_p+p}, (i) the
maximum energy a numerical total cross section
$\sigma_\mathrm{tot}^\mathrm{KK}(\sqrt{s_\mathrm{max}})$ can
achieve, depends on the minimum $x$ PDFs $f(x, Q)$ can give. For
CTEQ6.6 series, the believable maximum energy for the KK first-exist
state $M_1^\mathrm{KK}$ is only $7 \times 10^{15}\,\mathrm{eV}$,
which is insufficient for astrophysical purpose. Moreover, (ii) the
contributions of multi-KK producting processes lack of estimations
in our cross section calculations. Na\"{i}vely, one may want to
continue $\sigma_\mathrm{tot}^\mathrm{KK}(\sqrt{s})$ by fitting any
functions putting by hand, with our low energy numerical results
calculated above in Fig.~\ref{(fig)_cross_section}. However, if we
assume that $\sigma_\mathrm{tot}^\mathrm{KK}$ is dominated by some
special subprocess (as overpowered $g + g \rightarrow g_1^\ast +
g_1^\ast$ shows in Fig.~\ref{(fig)_cross_section} in our cases), it
may globally be inconsistent with the (axiomatic) Froissart-Martin
bound~\footnote{Of course, the Froissart-Martin bound need not to be
obeyed for a KK mass tower. Because its assumptions include that all
masses are equal to the unit of mass when the energy variables go to
infinity, it cannot have new states coming out when $\sqrt{s}$
increases. Spin structure is not a serious problem. Although
Froissart only derived that bound for scalar
particles~\cite{Froissart:1961ux}, we can imagine it is also
applicable for $2 \rightarrow 2$ quarks/gluons. Yet we can not
derive it out, just for a lack of some systematical ways parallel to
Mandelstam representation.} $\sigma_\mathrm{tot} \leq C \ln^2 s$ for
sufficiently large $s$~\cite{Froissart:1961ux}. Of course,
everything will be fine, assuming that Froissart-Martin bound is far
from being saturated.

\begin{figure}[ht]
\begin{center}
\includegraphics[angle=0,width=8cm]{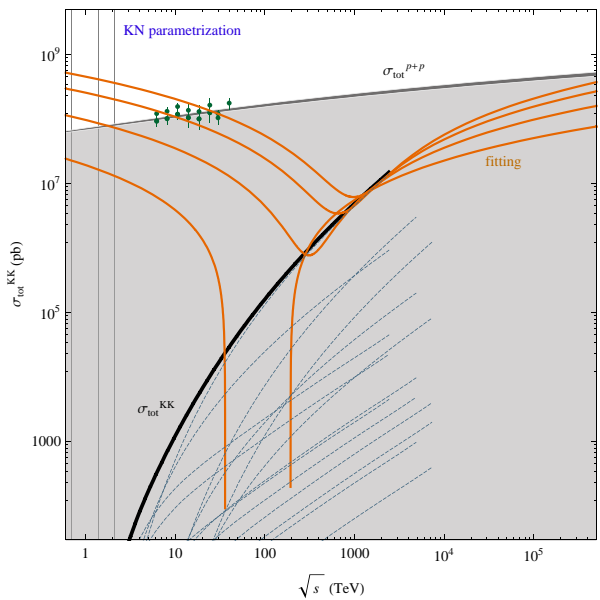}
\end{center} \caption{Fitting of dominate subprocess $g + g
\rightarrow g_1^\ast + g_1^\ast$ by the KN parametrization.
$\sigma_\mathrm{tot}^{p+p}$ is fitted in~\cite{Avila:2002tk} also by
the same parametrization, using the data of accelerator experiments,
Akeno, Fly's Eye, Nikolaev and
GSY~\cite{Baltrusaitis:1984ka,Gaisser:1986haa,Honda:1992kv,Nikolaev:1993mc}.
The data points in the top left corner are calculated from
$\sigma_\mathrm{tot}^{p+air}$ in air shower experiments. The thin
dashing lines are cross sections for subprocesses, as in
Fig.~\ref{(fig)_cross_section}, and
$\sigma_\mathrm{tot}^\mathrm{KK}$ is shown in a thick black line.
Because we can not in fact do a beautiful global fitting of
$\sigma^\mathrm{KK} (g + g \rightarrow g_1^\ast + g_1^\ast)$, as
shown in the figure, we fit it emphasizing particularly on (i) a
smoother derivative at the endpoint, or (ii) a preferable global
fitting. We will always use the first strategy in the follow-up
calculations in this context.} \label{(fig)_KN_para}
\end{figure}

\begin{figure}[ht]
\begin{center}
\includegraphics[angle=0,width=8cm]{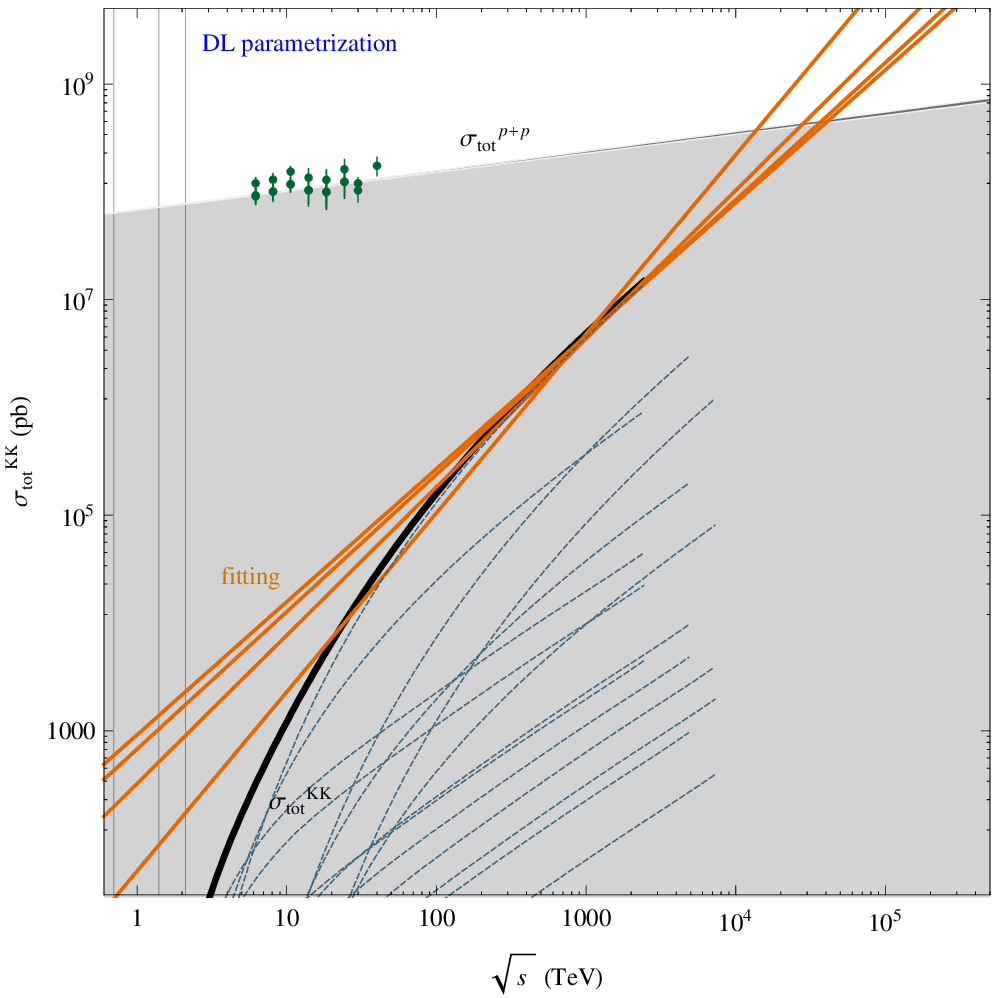}
\end{center} \caption{Sameline styles have been used as in
Fig.~\ref{(fig)_KN_para}, but otherwise the DL parametrization.
Because of some technical properties of Eq.~(\ref{(eq)_DL_para})
which cannot make it really looking like $\sigma^\mathrm{KK} (g + g
\rightarrow g_1^\ast + g_1^\ast)$ (see the context for detail), in
fact we only fit the cross section by $\sigma(\sqrt{s}) = D
s^\epsilon$. \label{(fig)_DL_para}}
\end{figure}

The high energy behavior of $p + p$ total cross section is
questioned in length~\cite{Block:1984ru}. In this context,
continuation of $\sigma_\mathrm{tot}^\mathrm{KK}(\sqrt{s})$ to
higher energy will be done by the Kang-Nicolescu (KN)
parametrization~\cite{Kang:1974gt}
\begin{equation}
    \sigma_\mathrm{tot}^{(2 \rightarrow 2)} = A + B \log s + C \log^2 s
    \label{(eq)_KN_para}
\end{equation}
and the Donnachie-Landshoff (DL) parametrization (Regge
fits)~\cite{Donnachie:1992ny}
\begin{equation}
    \sigma_\mathrm{tot}^{(2 \rightarrow 2)} = D s^\epsilon + E s^{-\delta} \mbox{.}
    \label{(eq)_DL_para}
\end{equation}
We choose them instead of arbitrary parametrizations simply because
(i) some unitarity constraints or physical backgrounds have already
been considered in their derivations, which may make them more
reasonable. The other reason is that (ii) the experimental result of
$p + p$ cross section for \emph{all} subprocesses
$\sigma_\mathrm{tot}^{p+p}$ estimated from proton-air inelastic
cross section
$\sigma_\mathrm{tot}^{p+air}$~\cite{Baltrusaitis:1984ka,Gaisser:1986haa,Durand:1987yv,Durand:1988cr,Honda:1992kv,Nikolaev:1993mc,Engel:1998pw,Avila:2002tk}
up to $\sqrt{s} \leq 10^{14}\,\mathrm{eV}$ in cascade processes, has
also been widely fitted by such kind of
parametrizations~\footnote{Extra subprocesses of $p + p$ cross
section from new physics beyond the SM, will hardly affect the
applicability of old fitting (in the air shower experimental
detectable region up to $10^{14}\,\mathrm{eV}$), because their
branching radios are always neglectable. When the energy becomes
higher, they may become more and more important, as shown in
Fig.~\ref{(fig)_KN_para} and \ref{(fig)_DL_para}. However, we
neglect this secondary correction in this context, simply because
the large theoretical uncertainties.}. Because
$\sigma_\mathrm{tot}^\mathrm{KK} / \sigma_\mathrm{tot}^{p+p}$ is an
important parameter in latter calculations, it will be convenient
for us to compute them from the same kind of parametrizations.

The continuations of dominate subprocess $g + g \rightarrow g_1^\ast
+ g_1^\ast$ by the KN and DL parametrizations, are shown in
Fig.~\ref{(fig)_KN_para} and \ref{(fig)_DL_para}. However, some
qualitative analyses show that, usually Eq.~(\ref{(eq)_DL_para}) can
only construct \emph{mild} concave functions in log-log diagram,
just as $\sigma_\mathrm{tot}^{p+p}$ in $\mathrm{TeV}$ regions. The
reason is that when both $D$ and $E > 0$, the two terms in
Eq.~(\ref{(eq)_DL_para}) are two asymptotes with a plus sign in
between. Eq.~(\ref{(eq)_DL_para}) can also construct convex
functions in a log-log diagram when $D \cdot E < 0$. That is, when
the negative term becomes dominate, the curve will drop rapidly,
which is not consistent with our purpose. Hence we simply fit
$\sigma(\sqrt{s}) = D s^\epsilon$ in Fig.~\ref{(fig)_DL_para}, with
the physical meaning of retaining the term analogical to pomeron
exchange but dropping out the term analogical to $\rho$, $\omega$,
$f$ and $a$ exchanges. We notice that both the KN and DL
parametrizations cannot do well to fit $\sigma^\mathrm{KK} (g + g
\rightarrow g_1^\ast + g_1^\ast)$, which hints us that $p + p
\rightarrow \mathrm{(the\,KK\,states)}$ are tremendously different
from $p + p \rightarrow \mathrm{(the\,(SM\,hadrons)}$, hence
appropriate parametrizations with reasonable physical background
need to be explored. However, in this context, we use only our rough
continuation for estimations below.

\subsection{Collision Producing\label{(subsec)_Collision_Producting}}

The KK bosons and fermions can be produced by $p + p$ collisions for
high energy protons accelerated in the Fermi
mechanism~\cite{Fermi:1949ee,Fermi:1954}, or a special lineage of it
as the diffusive shock acceleration
model~\cite{Axford:1977,Bell:1978zc,Bell:1978fj,Blandford:1978ky,Kirk:1987,Bednarz:1998pi,Keshet:2004ch}
(see Ref.~\cite{Drury:1983,Kirk:1999km} for a review), with the
cross section $\sigma_\mathrm{tot}^\mathrm{KK}$ already been
calculated in \S\ref{(subsec)_p+p} and
\S\ref{(subsec)_High_Energy_Behavior}. We want to know how many of
them can actually be produced in such environments.

To leave the pertinent computations for specific astrophysical
sources to the later publications, and to give a more universal
applicable estimation, we discuss by imitating the same assumptions
in deriving the Waxman-Bahcall (WB) bound as in
Ref.~\cite{Waxman:1997ti,Bahcall:1999yr}. Let $n_\mathrm{p}$ be
the number density and $E_\mathrm{p}$ the energy of protons, the
assumptions are as follows: (i) A percentage~\footnote{The reason is
that protons may need a threshold velocity $v_\mathrm{min}$ to be
actually accelerated. For example, in the diffuse shock model, the
needed energy is a few times of $m_\mathrm{p} v_\mathrm{s}^2 / 2$
for protons to have larger gyroradii than the shock
thickness~\cite{Bell:1978fj}.} of $\eta$ of the protons have an
injection spectrum of $d n_\mathrm{p} / d E_\mathrm{p} = K \cdot
E_\mathrm{p}^{-\alpha}$ isotropically distributed from their rest
mass energy $m_\mathrm{p} = 938\,\mathrm{MeV} \sim
10^9\,\mathrm{eV}$ to some cutoff energy $E_\mathrm{p,\,max}$ in the
source comoving frame, while a percentage of $(1 - \eta)$ of them
stay at rest in that frame. In this paper, we will assume $\eta = 1$
for simplicity. The generalization of our result to $\eta \neq 1$
cases are straightforward. The power law spectrum in energy is a
natural deduction of the Fermi mechanism~\cite{Fermi:1949ee} or
diffusive shock model~\cite{Drury:1983}. However, for diffusive
shock model, the existing derivations may need the assumption of
collisionless or scattering only elastically with infinitely massive
objects here and there, which is unfavorable for our $p + p$
producing. Hence, a discussion of energy spectrum in a shock with
influential inelastic scattering may be needed for our purpose. (ii)
Sources are in the critical optical depth $\tau_\mathrm{pp} = 1$,
and the optical depth is independent of protonic energy. When
deriving the WB bound as an upper limit, optically thin sources to
$p + p$ and $p + \gamma$ reactions have been assumed. Hence, the
meaning of these assumptions is that the WB bound has been
saturated. When estimating the upper bound of neutrinos, $p +
\gamma$ interactions may be more effective than $p + p$
interactions, thus they are given more attention. Producing of the
KK particles by $p + \gamma$ reactions is also interesting (see also
discussions at the beginning of \S\ref{(sec)_Producing}); however,
it is beyond the scope of this paper, by the lack of ready-made
calculations about amplitude-squared and cross section. Here, we
only consider $p + p$ interacting cases. Detailed discussions about
the competition of $p + p$ and $p + \gamma$ reactions to produce
both neutrinos and the KK particles are given in
\S\ref{(sec)_neutrino_detectors}; however, a wiser way is to discuss
them in association with specific astrophysical
scenarios~\cite{Paczynski:1994uv,Razzaque:2002kb,Razzaque:2003uw,Granot:2002qz}.
Hidden sources with optical depth $\tau_\mathrm{pp} \gg
1$~\cite{Razzaque:2003uv,Razzaque:2004yv} are very tempting for our
purpose, because they can produce more KK particles by collisions.
However, a big problem is how they can be accelerated and get away
from the source successfully. We will discuss this situation
together with some details of acceleration mechanism in
\S\ref{(subsec)_optical_depth}, comparing it with neutrino
observations in \S\ref{(sec)_neutrino_detectors}. Discussions
linking to more realistic astrophysical environments are left to
later publications.

\begin{figure}[ht]
\begin{center}
\includegraphics[angle=0,width=8cm]{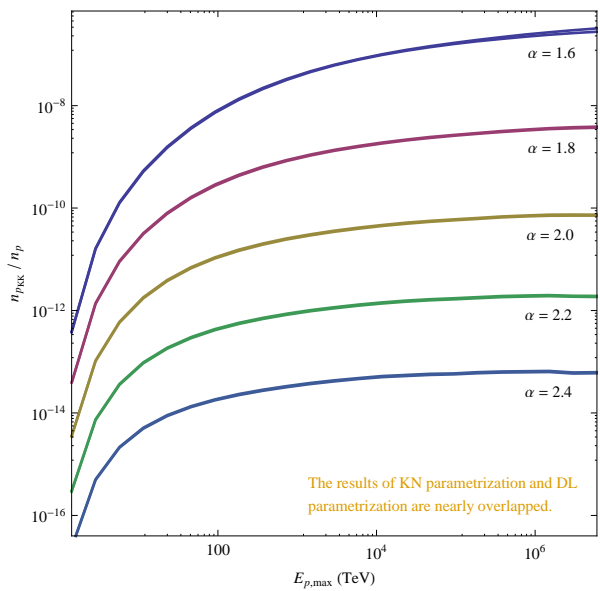}
\end{center}
\caption{The ratio $n_\mathrm{p_\mathrm{KK}} / n_\mathrm{p}$ after
one time of $p + p$ collision for each proton in a $E^{-\alpha}$
injection spectrum, depends on $E_\mathrm{p,\,max}$ and $\alpha$.
The results by the KN and DL parametrization are quite similar, thus
we cannot distinguish them in this figure. }
\label{(fig)_KK_production_Fermi}
\end{figure}

If we assume that normal $p + p$ interactions which dominate
$\sigma_\mathrm{tot}^{p+p}$ do not change the number of
protons~\footnote{It is reasonable because of baryon number
conservation. Other final states are either unstable (like strange
particles) or have small branching ratios (like the KK excited
states).}, after one time of collision, the number of the KK protons
in proportion to normal protons should be
\begin{equation}
\begin{split}
    \frac{n_\mathrm{p_\mathrm{KK}}}{n_\mathrm{p}} = &K^2\,
        \int_{m_\mathrm{p}}^{E_\mathrm{p,\,max}} E_1^{-\alpha} d E_1
        \int_{m_\mathrm{p}}^{E_\mathrm{p,\,max}} E_2^{-\alpha} d E_2
        \times \frac{2 \sigma_\mathrm{tot}^\mathrm{KK} (\sqrt{2 E_1 E_2})}{\sigma_\mathrm{tot}^{p+p} (\sqrt{2 E_1 E_2})}
        \mbox{,}
\end{split}
\end{equation}
in which we choose $\sqrt{s} = \sqrt{2 E_1 E_2}$ to neglect the
effect of rest mass $m_\mathrm{p}$ and simply choose $\theta =
\pi/2$. The resultant $n_\mathrm{p_\mathrm{KK}} / n_\mathrm{p}$
depending on $E_\mathrm{p,\,max}$ and $\alpha$ is shown in
Fig.~\ref{(fig)_KK_production_Fermi}. The cutoff energy
$E_\mathrm{p,\,max}$ in the source comoving frame can be calculated
by comparing the size of the shock and the duration of an
acceleration cycle, or acceleration gains and synchrotron losses of
energy (see Eq.~(\ref{(eq)_maximum_from_size}) and
(\ref{(eq)_maximum_from_synchrotron}) for details). Sources of
UHECRs should have $E_\mathrm{p,\,max} \geq 10^{18}\,\mathrm{eV}$ in
the shock comoving frame, because of the reason that we have already
seen a couple of UHECR events with energy $> 3 \times
10^{20}\,\mathrm{eV}$~\cite{Takeda:2002at,AbuZayyad:2002sf} in the
observer's frame. Even if neglecting the energy losses by the
Greisen-Zatsepin-Kuzmin (GZK)
effect~\cite{Greisen:1966jv,Zatsepin:1966jv} when propagating, and
dividing a Lorentz factor $\Gamma \sim 300$ of the source comoving
frame, we still need at least $E_\mathrm{p,\,max} =
10^{18}\,\mathrm{eV}$. We may prefer $\alpha = 3 v_\mathrm{d} /
(v_\mathrm{u} - v_\mathrm{d})$ for the nonrelativistic shock
cases~\cite{Drury:1983}, $\alpha = (3 \beta_\mathrm{u} - 2
\beta_\mathrm{u} \beta_\mathrm{d}^2 +
\beta_\mathrm{d}^3)/(\beta_\mathrm{u} - \beta_\mathrm{d}) - 2
\rightarrow 38/9 - 2 \simeq 2.2$ for ultra-relativistic shock
limit~\cite{Keshet:2004ch}, where $v_\mathrm{u}$
($\beta_\mathrm{u}$) and $v_\mathrm{d}$ ($\beta_\mathrm{d}$) the
upstream and downstream velocities ($\beta$ factors) in the shock
frame, $\alpha \leq 2$ by some statistical reasons of
sources~\cite{Waxman:1995vg}, or $\alpha = 2$ to be consistent with
the assumption of the WB bound~\cite{Waxman:1997ti,Bahcall:1999yr}.
We see that the proportion $n_\mathrm{p_\mathrm{KK}} / n_\mathrm{p}$
is not sensitive to either $E_\mathrm{p,\,max}$ (for a large enough
$E_\mathrm{p,\,max}$, which is easy to achieve for an actual
astrophysical environment) or the parametrization strategies we
choose, but sensitive to the spectral index $\alpha$.

\section{Accelerating and Propagating of the KK Protons\label{(sec)_accelerating_propagating}}

For the reason that the KK protons take charges, and may have a
not-very-short lifetime (see discussions in
\S\ref{(sec)_introduction}) before decaying to the LKP $\gamma_1$,
they can also be accelerated by the same mechanism similar to normal
protons. The acceleration process is never bothered by the KK
cascade decay. After being accelerated, they may also propagate
through the space for a considerable distance before decaying to
$\gamma_1$. If the decay processes are forbidden or suppressed
sufficiently by other reasons, they may propagate as far as banging
into the earth. Hence in this section, we discuss accelerating and
propagating properties of the KK protons, and also propagating
property of decay intermediate and final states.

\subsection{The Fermi Mechanism and Diffusive Shock Model}

While discussing the accelerated properties and bounds of an extreme
relativistic particle, some time scales are very important. We will
discuss them in the framework of the \emph{original} Fermi
acceleration system~\cite{Fermi:1949ee,Fermi:1954}, that is,
particles are randomly accelerated by the electromagnetic
turbulence. However, because of the deep connection between the
Fermi mechanism and diffusive shock model, some of the results are
also suitable for the latter. We will calculate these scales in the
source comoving frame (for the turbulence behind the shock, it is
just the shock comoving frame). To transform to the observer's frame
when the source itself is relativistic, both the size of the shock
$R$ and the energy of the particle $E$ should multiply the Lorentz
factor $\Gamma$, e.g., $R \rightarrow R/\Gamma$ and $E \rightarrow
E/\Gamma$.

The time scales are as follows: (i) The mean escape time
$t_\mathrm{E} = R / c$ measures the time scale to get through the
accelerating region. (ii) The Fermi acceleration time $t_\mathrm{A}
= \eta R_\mathrm{L} / c \mathbf{\beta}^2$~\cite{Hillas:1985is}
measures how quickly the particle gains energy, where $R_\mathrm{L}
= E / e B$ is the Larmor radius, $B$ is the magnetic field strength,
$\mathbf{\beta}$ is the Alfv\'{e}n velocity, and $\eta \sim
8/3-50/3$ is a factor determined by the turbulent system. (iii)
Synchrotron timescale $t_\mathrm{sy} = (6 \pi m^4 c^3 /
\sigma_\mathrm{T} m_\mathrm{e}^2) E^{-1} B^{-2}$ measures the energy
losses of synchrotron radiation, where $\sigma_\mathrm{T} = 8 \pi
r_\mathrm{e}^2 / 3$ is the Thomson cross section, $m = m_\mathrm{p}$
or $m = m_\mathrm{KK} \simeq M_n^\mathrm{KK}$ is the mass for both
normal or the KK protons respectively, and $M_n^\mathrm{KK}$ is the
KK excited quark mass. Successful acceleration needs both
$t_\mathrm{E} \geq t_\mathrm{A}$ and $t_\mathrm{sy} \geq
t_\mathrm{A}$, thus giving two limits of maximum acceleration energy
\begin{equation}
    E_\mathrm{max}^{(1)} \simeq \frac{e B R \mathbf{\beta}^2}{\eta}
    \label{(eq)_maximum_from_size}
\end{equation}
and
\begin{equation}
    E_\mathrm{max}^{(2)} \simeq \frac{m^2 c^2 \mathbf{\beta}}{m_\mathrm{e}}
        \sqrt{\frac{6 \pi e}{\eta \sigma_\mathrm{T} B}} \mbox{.}
    \label{(eq)_maximum_from_synchrotron}
\end{equation}
Because of the same charges but different masses normal and the KK
protons take, $E_\mathrm{max}^{(1)}$ cannot distinguish them but
$E_\mathrm{max}^{(2)}$ can. Hence, to find out the KK protons by the
higher energy they can achieve, we need $E_\mathrm{max}^{(1)} >
E_\mathrm{max}^{(2)}$ or the source to be sufficiently large, i.e.,
\begin{equation}
    r > \frac{\eta}{e B \mathbf{\beta}} \frac{m^2 c^2}{m_\mathrm{e}}
        \sqrt{\frac{6 \pi e}{\eta \sigma_\mathrm{T} B}} \mbox{.}
\end{equation}
To emphasize that up to a constant, the constraint $t_\mathrm{E}
\geq t_\mathrm{A}$ is equivalent to the assumption $R >
R_\mathrm{L}$, which is a universal requirement for accelerations
correlated to magnetic fields.

Notice that the formation of the cosmic-ray spectrum depends and
only depends on (i) the average gain in energy per acceleration
event, and (ii) the acceleration events a particle may suffer. For
the original Fermi mechanism~\cite{Fermi:1949ee}, the gain of energy
is $\delta E = (v/c)^2 E$ per collision in average (hence the energy
should enhance exponentially), where $v$ is the velocity of the
reflecting obstacles (e.g., the electromagnetic turbulence). Hence,
the spectrum index $\alpha = 1 + (c/v)^2 \Delta t_\mathrm{ela} /
\Delta t_\mathrm{inela}$ depends on the duration $\Delta
t_\mathrm{ela}$ between elastic scattering (with infinite massive
reflecting obstacles) and the duration $\Delta t_\mathrm{inela}$ to
break down the energetic particle. For the diffusive shock model in
non-relativistic shock wave cases~\cite{Kirk:1999km}, the average
momentum (hence energy for relativistic particles) gain for
isotropic distributed particles is $\delta p = 4 p (v_\mathrm{u} -
v_\mathrm{d}) / 3 v_\mathrm{p}$, and the probability of escape per
shock crossing is $4 v_\mathrm{d} / v_\mathrm{p}$, where
$v_\mathrm{p}$ is the velocity of accelerated particles. To
generalize this argument to relativistic shock wave cases, isotropic
distributions of particles are no longer applicable, thus an
intuitionistic derivation is absent. However, we may assume that the
qualitative phenomena are similar. Assuming that the total inelastic
cross section for the KK protons is similar to~\footnote{If the
cross section for the KK protons is smaller than that of protons, we
will have the more energetic KK cosmic-rays to be observed.} that of
protons, we have $\Delta t_\mathrm{inela,\,KK} \sim \Delta
t_\mathrm{inela,\,p}$, thus both (i) and (ii) are independent of
particles' properties. The only difference between protons and the
KK protons is the distinctness of their initial energy (rest mass).
Hence, we have reasons to believe that the energy spectrum of the KK
protons should shift to higher energy than protons by an amount of
$m_\mathrm{KK} / m_\mathrm{p}$.

\begin{figure}[ht]
\begin{center}
\includegraphics[angle=0,width=8cm]{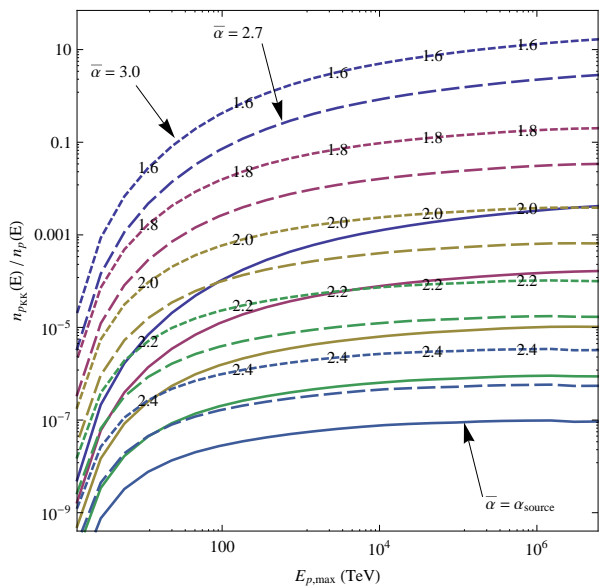}
\end{center}
\caption{$n_\mathrm{p_{KK}} / n_\mathrm{p}$ for fixed $E$ when
$\bar{\alpha} = 2.7$, $3.0$ (for diffuse flux) and
$\alpha_\mathrm{source}$ (for special astrophysical sources). It is
not too small to be identified.}
\label{(fig)_KK_modified_energy_shift}
\end{figure}

The resultant $n_\mathrm{p_{KK}} / n_\mathrm{p}$ for interzone of
fixed observed $E$ is shown in
Fig.~\ref{(fig)_KK_modified_energy_shift}, which is independent of
$E$ for $m_\mathrm{KK} < E < E_\mathrm{max}$. Notice that
$n_\mathrm{p_{KK}}$ is enhanced by a factor of $(m_\mathrm{KK} /
m_\mathrm{p})^{\bar{\alpha}}$ by the spectrum shift, where
$\bar{\alpha}$ is the index of the spectrum. We choose $\bar{\alpha}
= 2.7$ or $3.0$ for the overall cosmic-ray spectrum observed below
and above the ``knee'' energy $E_\mathrm{knee} \simeq
10^{15}\,\mathrm{eV}$, and $\bar{\alpha} = \alpha$ for some special
sources. For reasonable astrophysical sources, $n_\mathrm{p_{KK}} /
n_\mathrm{p}$ may be not too small a number, to that we have
opportunities to discover them by air shower identification.

\subsection{$p + p$ Optical Depth Reexamined\label{(subsec)_optical_depth}}

Taking no account of the details of especial astrophysical sources,
the $p + p$ optical depth $\tau_\mathrm{pp}$ (which is assumed to be
$1$ in \S\ref{(subsec)_Collision_Producting} and subsequently
quantitative estimations) can be discussed by focusing on special
acceleration mechanisms.

In the case of the \emph{original} Fermi
mechanism~\cite{Fermi:1949ee,Fermi:1954}, $\tau_\mathrm{pp} \geq 1$
can be achieved by a sufficiently long existent duration of
acceleration sources. However, high energy protons can also be
broken down by $p + \gamma$ interaction or other processes, so
several times of accelerations may be needed.

In the case of diffusive shock model, things are a little more
troublesome. As we have already indicated in
\S\ref{(subsec)_Collision_Producting}, the derivation of power law
spectrum itself needs the assumption of collisionless. In the
mainstream tactics for particles to cross back and forth the shock
front, the downstream particles are scattered by strong magnetic
turbulence behind the shock, and the upstream particles are
scattered by the self-excited Alfv\'{e}n waves~\cite{Drury:1983}.
The $p + p$ collisional disintegration has to be competitive with
particle missing in downstream scattering, which makes a constraint
for its feasible parameter space (because we already know some
restrictions from spectrum index, etc). However, we should not be
too serious in that problem, because until now there is still no
consistent computation in diffusive shock model to generate the kind
of waves, scatterings and acceleration
phenomena~\cite{Keshet:2004ch}. Scattering with analogously
energetic particles (rather than with infinite massive objects) is
also a good way to change direction but avoid loosing too much
energy, especially in the upstream regions where self-excited
Alfv\'{e}n waves themselves are in trouble.

\subsection{The Modified GZK Cutoff\label{(subsec)_GZK_cutoff}}

For the case that (i) the KK protons have too long a lifetime to
decay while propagating, when discussing the propagating of protons
versus them, the most (or the only) important issue is the
modification of the GZK
cutoff~\cite{Greisen:1966jv,Zatsepin:1966jv}. Other issues while
propagating, such as $e^{+} e^{-}$ pair production, cannot change
tremendously the property of energy spectrum. The GZK cutoff, which
is the threshold energy for protons to interact with the cosmic
microwave background (CMB) photons to produce pions, as
\begin{equation}
    E_{\mathrm{GZK}} \simeq \frac{m_\pi^2 + 2 m m_\pi}{4 E_\mathrm{CMB}}
    \simeq
    6 \times 10^{19}\,\mathrm{eV} \left(\frac{m}{m_\mathrm{p}}\right)
    \label{(eq)_GZK_threshold}
\end{equation}
for $m \gg m_\pi$, is enhanced for the KK protons by $m_\mathrm{KK}
/ m_\mathrm{p}$ times. We have assumed that the quarks and gluons in
pions are all the KK zero-mode (because the threshold energy to
produce the KK excited pion is another $m_\mathrm{KK} / m_\pi$ times
higher), and the cross section for $\gamma_\mathrm{CMB} +
p_\mathrm{KK} \rightarrow p_\mathrm{KK} (n_\mathrm{KK}) + \pi^0
(\pi^+)$ is still large enough to make a cutoff. The second
assumption is irrelevant because the more observable KK excited
cosmic-rays lend themselves to discovery; however, the number of
particles is reduced tremendously for higher energies so we have
little opportunities to really meet one.

For the opposite case that (ii) the KK protons have already decayed
to the LKP $\gamma_1$ (for other cases, including the decay
intermediate states, the discussions are analogous; thus we will not
discuss this further in this paper), they may also interact with CMB
photons, as $\gamma_\mathrm{CMB} + \gamma_1 \rightarrow X$. However,
we do not know what $X$ really is, whether the cross section is
sufficiently large to make a spectral cutoff, or what threshold
energy these processes correspond to. Because the LKP $\gamma_1$ is
an elementary particle (rather than the KK protons which include
normal quarks/glouns), its cross sections may be much different from
that of normal ones. If the process $\gamma_\mathrm{CMB} + \gamma_1
\rightarrow \gamma_1 + \pi^0$ is sufficiently important, the
threshold has an enhancement of $m_\mathrm{KK} / m_\pi$ thus is much
larger than the GZK cutoff energy, similar to
Eq.~(\ref{(eq)_GZK_threshold}).

\section{Air Shower Identification\label{(sec)_air_shower_identification}}

If we can distinguish air shower events in which primary particles
are protons/ions or the KK particles (the KK protons, the LKP
$\gamma_1$ or other intermediate KK excited particles which bang
into the earth), we are capable of giving a tighter bound on KK
contamination. Otherwise, we can only constrain it when the KK
particles dominate the mass composition.

The mass composition of cosmic-rays is rudimentary, partly because
cascade processes with the same initial condition are not identical
with each other, partly because of our insufficient knowledge of
parton distribution, thus different cascade models (e.g.
QGSJET~\cite{Kalmykov:1997te} and SIBYLL~\cite{Fletcher:1994bd})
cannot make consistent predictions with each other. An incomplete
set of parameters to discriminate different compositions of
cosmic-rays are as follows: (i) the elongation rate/shower maxima
$X_\mathrm{max}$~\cite{Gaisser:1993ix}, when the energy of the
particle can be decided separately by fluorescence technique, (ii)
the magnitude of the fluctuation in depth of maximum
$X_\mathrm{max}$~\cite{Abbasi:2004nz}, (iii) the number of particles
$n_\mathrm{max}$ at $X_\mathrm{max}$, (iv) the speed of rise in
$n_\mathrm{max}$~\cite{Ambrosio:2005bs}, (v) fraction of muons in
the shower events, and some geometrically-base parameters like (vi)
lateral distribution functions (LDFs), (vii) the thickness of the
shower disk or (viii) the shower front curvature (see
Ref.~\cite{Watson:2004ew} for a review). Some of the diversities
rise by the simulation results of cascade
models~\cite{Kalmykov:1997te,Fletcher:1994bd}. Hence, to identify a
KK excited cosmic-ray event for our purpose, the PDFs for protons
take the KK charge and corresponding air shower simulations of (i)
$p_\mathrm{KK}$, (ii) $\gamma_1$ or (iii) other intermediate KK
excited particles are needed. However, they are beyond our scope in
this paper.

Na\"{i}vely, differences mainly rise for the reason that ions are
made up from constituent nucleon of relatively lower energy
(typically as (iii, iv)~\cite{Ambrosio:2005bs} and the Zatsepin
effect~\cite{Gerasimova:1960,MedinaTanco:1998ac}) cannot be used for
reference, so a careful filtration of the derived parameters from
above is needed. For some parameters, their values can only be
determined statistically (such as (i) in Fly's Eye data
processing~\cite{Gaisser:1993ix}) at present, and are thus not
applicable for our purpose.

For the case of (i) the KK protons, similar total cross sections
$\sigma_\mathrm{tot}$ for $p + p$ and $p + p_\mathrm{KK}$ are
expected, because $p_\mathrm{KK}$ has two/three normal valence
quarks (and uncountable sea quarks) sharing energy, and
$\sigma_\mathrm{tot} (p + p)$ is not supposed to increase
tremendously for larger center-of-mass energy $\sqrt{s}$ (see the
discussions for $p_\mathrm{KK}$ production in high energy $p + p$
collision in \S\ref{(sec)_Producing} for detail). Hence, the primary
KK protons should also trigger \emph{a priori} cascade processes.
Differences may rise from the mass hierarchy between $p$ and
$p_\mathrm{KK}$, or equivalently from the energy transformation
between normal and the KK quarks/gluons that should absolutely
happen by color confinement, thus making the differential cross
section $d \sigma (p+p) / d \Omega$ and $d \sigma (p+p_\mathrm{KK})
/ d \Omega$ dissimilar. That may make the first few collisions much
different; however, the phenomenology for the secondary showers of
cascade particles like $\pi^\pm$, $\pi^0$, $p$, $\bar{p}$, $e^\pm$
and $\gamma$ are the same. Some similar work in
Ref.~\cite{Anchordoqui:2007pn} and \cite{Anchordoqui:2008zz},
in which the cascade process of hadronized gluinos (analogous to our
KK protons with the KK quarks/gluons in them) bang into the earth is
simulated, indicate that the identification is possible; however,
other researchers are suspicious of their
result~\cite{Kopenkin:2008zz}. It will be favorable for us, if the
former result is correct, and can be naturalized to our KK proton
case. Observable air shower events for the KK electrons are also
expected, because it has turned on quantum electrodynamics (QED)
interaction, thus the cross section should not be very small.
However, the origin of the KK electrons is not discussed in this
context. As a result, we can only distinguish primary protons/ions
or the KK charged particles for their different cascade properties,
and quantitative analyses are needed for the identifiability.

For the case of identifying (ii) the LKP $\gamma_1$, things might be
a little easier. We have already had trustworthy methods to identify
photons from air shower data~\cite{Halzen:1994gy,Risse:2004mx},
because they interact with the geomagnetic field, thus starting the
cascade process much before that of protons. Hence, if $\gamma_1$
can also interact with geomagnetic field, we can identify them
easily. However, the quantitative calculations need detailed
properties of $\gamma_1$ interactions.

Exotic cosmic-ray events have already been regarded, for example in
Ref.~\cite{Chen:1997qx}. However, they may be particularly
interested as the weakly interacting low energy ones, which have
their first collisions inside our scientific equipments. For our
purpose, it is more pertinent to find UHE exotic air-shower events.
If they would happen, they may be explained by the UEDs scenario we
mention in this paper (or other new physics models), and their
absence can constrain the same theoretical models.

If we cannot identify the KK charged particles' cascade events from
normal protonic/ionic ones, we can only constrain it when the KK
particles dominate the mass composition. A cumbersome issue is that
energy measurement is also related to the assumption of primary
particles' composition. So that even if there have already been the
super-energetic KK particles (with energy largely exceeding the GZK
threshold) banging into the earth, we might not discover them by our
energy estimation. A wise way for energy measurements is to choose
some kind of calorimetric measurements (like fluorescence light
emissions~\cite{Linsley:1983ch,Song:1999wq}) which are relatively
model independent. Here in this paper, we assume that energy
spectrum can be determined without overall departure for
protons/ions or the KK particles. Thus we are capable of discovering
them when they dominate the mass composition, leading to some
identifiable phenomena such as a lack of the GZK cutoff, etc.

As shown in Fig.~\ref{(fig)_KK_modified_energy_shift}, if assuming
$\tau_\mathrm{pp} = 1$ and $\alpha = 2$, the resultant KK
contamination $n_\mathrm{p_{KK}}(E) / n_\mathrm{p}(E)$ may be as
large as $10^{-5}$ to $10^{-2}$. Hence if we can identify them
individually from protonic/ionic ones, it is not difficult to make a
discovery. However, if we can not in fact identify them from air
shower data, a larger $\tau_\mathrm{pp}$ may be needed.

\section{$\tau_\mathrm{pp}$ Constraints from High Energy Neutrino Detectors\label{(sec)_neutrino_detectors}}

The $p + p$ optical depth $\tau_\mathrm{pp}$ is a crucial parameter
for the applicability of our methodology; however, it should be
discussed in relation to special astrophysical sources. Quantitative
calculations are difficult because the models for UHE particle
origins themselves, such as gamma-ray bursts (GRBs), active galactic
nuclei (AGNs) or supernovae (SNe), are also full of dubious issues.

However, an universal estimation can be made from the observation of
neutrinos. The reason is that $p + p$ collisions can produce the KK
protons, but also have branching ratios to produce $\pi^\pm$, which
decay to neutrinos mainly in modes $\pi^\pm \rightarrow \mu^\pm +
\nu_\mu (\bar{\nu}_\mu) \rightarrow e^\pm + \nu_e (\bar{\nu}_e) +
\nu_\mu + \bar{\nu}_\mu$~\footnote{Other processes such as $K^\pm
\rightarrow e^\pm + \nu_e (\bar{\nu}_e) + \nu_\mu + \bar{\nu}_\mu$
may also be important for our purpose, because kaons loose less
energy before decaying into neutrinos, by their relatively larger
mass and shorter lifetime~\cite{Ando:2005xi}. However, we will not
discuss this further in this paper.}. Thus if we have known (or have
an upper limit of) neutrino flux in earth, we can use it to restrict
optical depth $\tau_\mathrm{pp}$ as well as the KK cosmic-ray flux.
Of course, the constraint can only be used as an upper limit,
because we do not know what percentage of neutrinos are in fact
produced by $p + \gamma$ interactions compared to $p + p$
interactions, or whether the KK particles can be accelerated
successfully, or whether they can leave the source as easily as
neutrinos do.

There are already some publications of $p + p$ collisions of the
Fermi accelerated protons in astrophysical
environments~\cite{Paczynski:1994uv,Razzaque:2002kb,Razzaque:2003uw,Granot:2002qz,Razzaque:2003uv,Razzaque:2004yv,Murase:2008sp}.
However, they are not of much use for our purpose, mainly because:
(i) Their calculations are always oversimplified, because $p + p$
interactions are always unimportant, compared to $p + \gamma$
interactions in specific astrophysical environments. (ii) Their
discussions do not include subprocesses opposite to $p + p
\rightarrow \pi + X$. (iii) Within some of their scenarios, one of
the reacting protons does not suffer Fermi acceleration. In our
scenario, it is efficient (and also favored) that both protons are
accelerated. (iv) They make few attempts to calculate neutrino
spectrum, or still assume $E_\nu \simeq 0.25 E_\pi \simeq 0.05
E_\mathrm{p}$ all the same. The latter simplification is reasonable
in the $p + \gamma$ cases, but not legitimate for $p + p$ cases,
because usually $E_\mathrm{p} \gg m_\mathrm{p} \gg E_\gamma$ makes
$\Gamma_\mathrm{p} \sim \Gamma_\pi \sim (E_\mathrm{p} +
E_\gamma)/\sqrt{s}$, thus energy can be shared roughly by mass
proportion. Although spectra of \emph{thermal} $p + p$ interaction
have already been discussed in Ref.~\cite{Dermer:1986}; however,
a careful calculation of energy spectrum of \emph{nonthermal} $p +
p$ collisions (especially in the case where both protons are
accelerated by the Fermi mechanism) is absent. Briefly, only when
both protons have roughly the same (direction and amplitude of the)
momenta, the simplification $E_\nu \simeq 0.05 E_\mathrm{p}$ can be
used; however, the spectrum of neutrinos should be suppressed by
$E^{-2\alpha}$ rather than $E^{-\alpha}$ in that case. Hence, we may
guess the energy spectrum of $p + p$ neutrinos is steeper than
$E^{-\alpha}$ in higher energies. Some numerical results for
specific astrophysical environments show that the energy spectrum
looks like $E^{-(\alpha+1)}$~\cite{Razzaque:2002kb}. We will give a
more detailed universal calculation in \S\ref{(subsec)_pp_spectrum}.

\subsection{$p + p \rightarrow \pi + X$ Cross Sections}

\begin{figure}[ht]
\begin{center}
\includegraphics[angle=0,width=8cm]{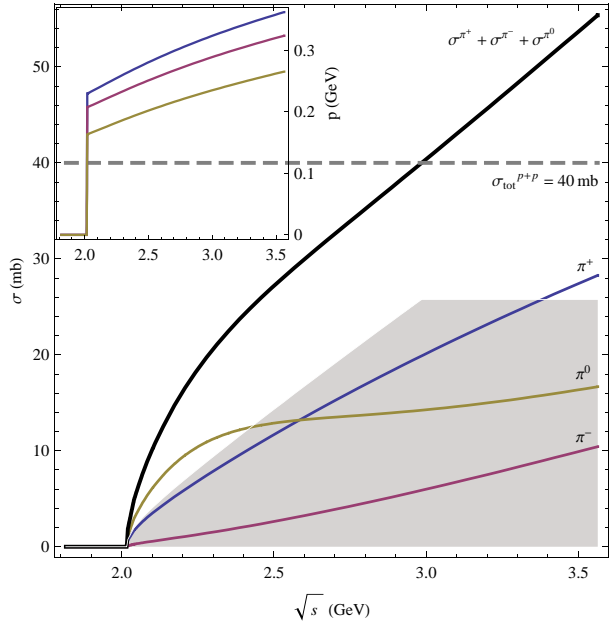}
\end{center}
\caption{Numerical result of total cross section $\sigma^\pi
(\sqrt{s})$ within Badhwar parameterization~\cite{Badhwar:1977zf}.
The dark region shows that the parameterization is only suitable for
$\sqrt{s} \leq 2.98\,\mathrm{GeV}$, thus we force-fix the proportion
$\sigma^{\pi^+} : \sigma^{\pi^-} : \sigma^{\pi^0}$ for large
$\sqrt{s}$. The weighting average $\langle p^\star_\pi \rangle
(\sqrt{s})$ is in the top left corner of this figure, with the same
abscissa as $\sigma^\pi (\sqrt{s})$.}
\label{(fig)_Badhwar_cross_section}
\end{figure}

The cross sections to produce $\pi^\pm$ in $p + p \rightarrow \pi +
X$ collisions have already been calculated
in~\cite{Carey:1974zd,Alper:1975jm,Badhwar:1977zf,Ellis:1977rz} (see
also~\cite{Blattnig:2000zf,Norbury:2007kf} for recent developments).
These parameterizations are mainly based on relatively low energy
terrestrial experiments, and focus on Lorentz-invariant differential
cross section (LIDCS) $E (d^3 \sigma / d^3 \mathbf{p})$ rather than
total cross section $\sigma_\mathrm{tot}$. Simple estimations show
that they may behave well near the center-of-mass threshold energy
$\sqrt{s} \geq 2 m_\mathrm{p} + m_\pi$ to produce pions, but
increase too quickly to be consistent with the total cross section
for all $p + p$ subprocesses
$\sigma_\mathrm{tot}^{p+p}$~\cite{Kang:1974gt,Donnachie:1992ny,Avila:2002tk}
we use in this context. Here, to give a consistent but not too
cumbersome estimation from an astrophysical (rather than collider
physical) viewpoint, we use Badhwar
parameterization~\cite{Badhwar:1977zf} when $\sigma^{\pi^+} +
\sigma^{\pi^-} + \sigma^{\pi^0} \leq \sigma_\mathrm{tot}^{p+p} \sim
40\,\mathrm{mb}$ (thus $\sqrt{s} \leq 2.98\,\mathrm{GeV}$), and
assume all $\sigma_\mathrm{tot}^{p+p}$ produce pions for larger
$\sqrt{s}$ with total cross section $\sigma_\mathrm{tot}^{p+p}$ for
a fixed proportion as the same as when $\sigma^{\pi^+} +
\sigma^{\pi^-} + \sigma^{\pi^0} = \sigma_\mathrm{tot}^{p+p}$. The
result is shown in Fig.~\ref{(fig)_Badhwar_cross_section}. In the
calculation, we assume (i) each $p + p$ interaction produces single
pion, and (ii) $m_\mathrm{X} \simeq 2
m_\mathrm{p}$~\cite{Nagamiya:1982kn} when evaluating the
$p_{\parallel,\,\mathrm{max}}$ in Feynman scaling variable
$x_\mathrm{F} = p_\parallel / p_{\parallel,\,\mathrm{max}}$. The
applicability of both assumptions must be considered carefully for
our purpose.

For a more accurate estimation of $p + p$ neutrino spectrum, the
LIDCS $E (d^3 \sigma / d^3 \mathbf{p})$ (which depend on $\sqrt{s}$,
the pion energy $E^\star_\pi$ and the scattering angle
$\theta^\star$ in the center-of-mass frame~\footnote{In this papaer,
we use superscript $\star$ to denote the center of mass frame, and
normal characters for the source comoving frame.}) should be used
directly to evaluate $\Gamma_\pi = \Gamma_\mathrm{cm}
\Gamma_\pi^\star (1 + \mathbf{v}_\mathrm{cm} \cdot
\mathbf{v}_\pi^\star / c^2)$. However, it pours oil on the flames in
the further complicated calculations. Hence, we integrate out
$E^\star_\pi$ and $\theta^\star$ by
\begin{equation}
\begin{split}
    \sigma^\pi (\sqrt{s}) =
    &2 \pi \int_0^\pi \sin\theta^\star d \theta^\star
        \int_0^\infty d p^\star_\pi
        \times \frac{p_\pi^{\star\,2}}{\sqrt{p_\pi^{\star\,2} + m_\pi^2}}
        \cdot E^\star_\pi \left( \frac{d^3 \sigma^\pi}{d^3 \mathbf{p^\star}_\pi} \right)
        \mbox{,}
\end{split}
\end{equation}
and memorize $p^\star_\pi$ thereby $\Gamma_\pi^\star = E^\star_\pi /
m_\pi = \sqrt{p^{\star\,2}_\pi + m_\pi^2} / m_\pi$ only by a
weighting average
\begin{equation}
    \langle p^\star_\pi \rangle (\sqrt{s})
        = \left. \int p^\star_\pi (\ldots) \right/ \int (\ldots)
\end{equation}
where $\int (\ldots) = \sigma^\pi (\sqrt{s})$. In addition, we
neglect the effect of $\mathbf{v}_\mathrm{cm} \cdot
\mathbf{v}_\pi^\star / c^2$. The result is in the top left corner of
Fig.~\ref{(fig)_Badhwar_cross_section}. We see that for $\sqrt{s}$
not too close to $2 m_\mathrm{p} + m_\pi$, it is always true that
$\sqrt{s} - 2 m_\mathrm{p} - m_\pi \gg \langle p^\star_\pi \rangle$.
However, the result is not suitable for larger center-of-mass
energy, and we lack a reasonable extension for $\sqrt{s} >
2.98\,\mathrm{GeV}$.

\subsection{Neutrino Spectrum\label{(subsec)_pp_spectrum}}

Hence, if we know the spectrum and optical depth of protons (within
some specific astrophysical scenarios), the cross sections for $p +
p$ to produce $\pi^\pm \rightarrow e^\pm + \nu_e (\bar{\nu}_e) +
\nu_\mu + \bar{\nu}_\mu$, we can calculate the spectrum of neutrinos
produced by $p + p$ collisions. However, the calculation is not
easily done. Even if decoupling the relationship between vectors
$\mathbf{v}_\mathrm{cm}$ and $\mathbf{v}_\pi^\star$, and using only
a weighting average momentum $\langle p^\star_\pi \rangle$ for some
specific $\sqrt{s}$, the pion distribution function should look as
complicated as
\begin{equation}
\begin{split}
    n_\pi(\Gamma_\pi) \sim\,&K^2\,
        \int_{m_\mathrm{p}}^{E_\mathrm{p,\,max}} E_1^{-\alpha} d E_1
        \int_{m_\mathrm{p}}^{E_\mathrm{p,\,max}} E_2^{-\alpha} d E_2\\
        &\times \frac{1}{2}\int_0^\pi \sin \theta d \theta
        \frac{\sigma^{\pi^+}(\sqrt{s}) + \sigma^{\pi^-}(\sqrt{s})}{\sigma_\mathrm{tot}^\mathrm{p+p}(\sqrt{s})}\\
        &\times \delta \left( \Gamma_\pi -
        \frac{\sqrt{\langle p^\star_\pi \rangle^2(\sqrt{s}) + m_\pi^2}}{m_\pi}
        \frac{E_1 + E_2}{\sqrt{s}}\right)
        \mbox{,}
\end{split}
\end{equation}
where $\sqrt{s}(E_1,E_2,\theta)$ is the center-of-mass energy in the
source comoving frame. To deal with the Delta function in an easier
way, we fixed $\langle p^\star_\pi \rangle(\sqrt{s}) \equiv
0.2\,\mathrm{GeV}$ and integrate out $\theta$ visually. However, our
simplifications may underestimate the harder part of the spectrum.
To estimate the absolute value of flux to normalize our calculation,
we use the same assumption as in deriving the WB
bound~\cite{Waxman:1997ti,Bahcall:1999yr}, that is, the cosmic-rays'
energy production rate in the nearby universe is $(E_\mathrm{CR}^2\
d \dot{n}_\mathrm{CR} / d E_\mathrm{CR})_{z = 0} =
10^{44}\,\mathrm{erg\,Mpc^{-3}\,yr^{-1}}$, or equally
$E_\mathrm{CR}^2\ d n_\mathrm{CR} / d E_\mathrm{CR} = 5.05 \times
10^{-8}\,\mathrm{GeV\,cm^{-2}\,s^{-1}\,sr^{-1}}$ (or $3$ times
larger if thinking over redshift evolution).

\begin{figure}[ht]
\begin{center}
\includegraphics[angle=0,width=8cm]{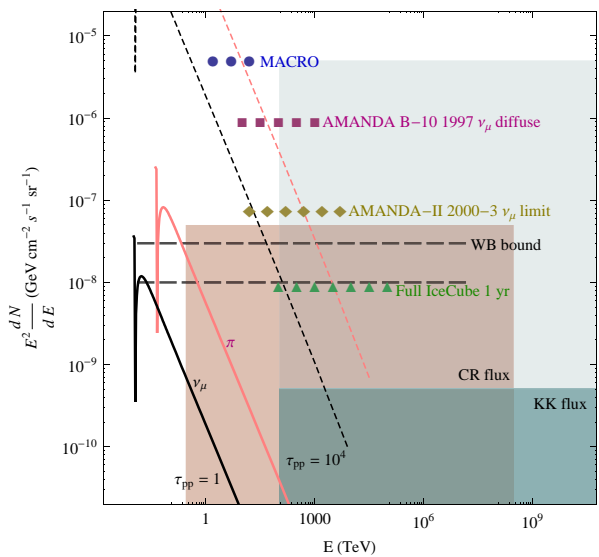}
\end{center}
\caption{$\nu_\mu$ flux from $p + p \rightarrow \pi + X$ collisions.
The shadow regions denote cosmic-ray flux, with the KK ones for
$\tau_\mathrm{pp} = 1$ and $\tau_\mathrm{pp} = 10^4$ respectively.
The detector information is cited from~\cite{IceCube:2007}. Current
neutrino detectors can only give constraints that $\tau_\mathrm{pp}
< 10^4$.} \label{(fig)_lots_of_bounds}
\end{figure}

The resultant spectra are in Fig.~\ref{(fig)_lots_of_bounds}. Our
simplified estimation of $\pi$ and $\nu_\mu$ spectra are consistent
with the numerical result $E^{-(\alpha+1)}$ in
Ref.~\cite{Razzaque:2002kb}. In our estimation, we have already
considered the effect of neutrino oscillation (detectors are only
sensitive $\nu_\mu$ flux). The WD bound can only be treated up to a
constant, because if using the same treatment as ours, they should
be much lower. The KK flux is estimated in a $\alpha = 2$,
$\bar{\alpha} = 3$ environment (thus we use $n_\mathrm{p_{KK}}(E) /
n_\mathrm{p}(E) \simeq 10^{-2}$), other choices of test parameters
can be read out directly from
Fig.~\ref{(fig)_KK_modified_energy_shift}. We see that
$\tau_\mathrm{pp}$ is not under strong constraints by neutrino
observations. However, when we have more advanced neutrino
detectors, they may give stronger constraints on $\tau_\mathrm{pp}$
hence the KK flux.

\subsection{Other Possible Constraints of $\tau_\mathrm{pp}$ besides $\nu_\mu$ Observations}

Another possible constraint of $\tau_\mathrm{pp}$ besides neutrino
flux observation is $\mathrm{GeV} - \mathrm{TeV}$ gamma-ray flux
observations. The reason is that $p + p$ collisions can not only
produce $\pi^\pm \rightarrow e^\pm + \nu_e (\bar{\nu}_e) + \nu_\mu +
\bar{\nu}_\mu$, but it can also produce $\pi^0 \rightarrow 2
\gamma$. Hence, the EGRET, GLAST or Swift telescope can give it a
constraint. Ref.~\cite{Anchordoqui:2004bd}, in which one
energetic and one rested proton collide to produce a
not-very-short-lived strongly interacting massive particles,
considered both neutrino and gamma-ray flux constraints. We lack of
discuss the latter one, because of the fact that both (i) to avoid
this paper to be too long (the discussion is very similar), and (ii)
to be different from energetic neutrinos which can only be produced
by violent or high $\Gamma_\mathrm{cm}$ collisions, energetic
gamma-rays can also be produced by the more mild synchrotron self
Compton (SSC) processes~\cite{Gupta:2007yb}. Hence the gamma-ray
constraint is not as na\"{i}ve to deal with as the neutrino ones, so
we leave it to later publications.

\section{Discussion and Outlook\label{(sec)_discussion_outlook}}

Phenomenologies link ambitious theoretical physical models to
reality, thus make physics a \emph{science}~\footnote{At least in
the philosophy of Karl Popper or previously.}. Terrestrial
experiments are one of the ordinary methods to constraint new
physics models; however, they have limited power because of our
finite energy sources on earth. Cosmology can open an extraordinary
window for new physics studies; however, Big Bang (and the extreme
physical environment it has) happened only once in our universe,
thus makes re-enactment impossible. Astrophysical constraints always
have larger scope than terrestrial experiments (because they do not
have the upper threshold of maximum achievable energy); however,
meaningful scenarios presently known are always only available for
weakly interacting light particles. The motivation of this paper is
to search for another way to construct new physics beyond the SM.

In this paper, we construct an astrophysical scenario to
(dis)confirm new physics for heavy particles beyond $\mathrm{TeV}$
energy scale. In our scenario, the KK protons are produced by $p +
p$ collisions in Fermi accelerated environments, and they themselves
are accelerated by the same environments. Because they may change
the compositions and proterties of cosmic-ray events, air shower
experiments can give a constraint to their properties.

To know whether our scenario can give meaningful constraints to UEDs
(and maybe other new physics models in later researches), we make
some quantitative estimations. We first investigate whether enough
KK excited states can be produced by $p + p$ collisions. We
calculate the overall KK cross sections by precalculated Feynman
rules and amplitude-squared, with also the CTEQ6.6m proton PDFs.
Subprocesses $g + g \rightarrow g_n^\ast + g_n^\ast$ and $q + q
\rightarrow q_n^\ast + q_n^\circ$ may be most important in quark
level. Because of color confinement, the KK excited quarks and
gluons should form the KK excited protons. We then calculate what
percentage of the KK protons can be produced by an isotropic and
power law distributed proton spectrum. For the spectral index
$\alpha = 2.0$, $n_\mathrm{p_\mathrm{KK}} / n_\mathrm{p} \sim
10^{-10}$ as an overall contamination after one time of $p + p$
collision. However, when considering that the KK protons are also
accelerated by the Fermi mechanism, the scene is much different. It
is reasonable to believe that the energy spectrum of the KK protons
should shift to higher energy than protons by an amount of
$m_\mathrm{KK} / m_\mathrm{p}$, hence for some fixed energy $E$, the
KK states should contaminate $10^{-5}$ (for some special
astrophysical sources with the spectral index $\alpha = 2.0$) to
$10^{-2}$ (for diffuse flux) of cosmic-ray events. So, if we can
identify them from other cosmic-ray particles from air shower data,
our method is capable of giving meaningful constraints. We notice
that the GZK cutoff energy also shifts by a factor of $m_\mathrm{KK}
/ m_\mathrm{p}$ to higher energy, if it still exists. Hence,
observations of cosmic-ray particles with energy much above the GZK
cutoff, are given a reasonable explanation by the KK particles;
however, the possibilities of this kind of observation is really
small, even if the GZK cutoff does not exist. We also investigate
the possibilities to identify the KK cosmic-ray events by air shower
data. The investigation is still superficial, because quantitative
simulations are needed; however, the LKP $\gamma_1$ may be easy to
identify, because they may interact with geomagnetic fields just
like normal photons, thus make the air shower tomography much
different from that of protons. Finally, we calculate the possible
constraints of the KK cosmic-ray flux from neutrino detectors;
however, the constraints are very loose for current scientific
equipments to affect our former estimations.

It is appropriate to regard our calculations (outlined in this
paper) as an ``existence proof'' for this kind of methodology. In
fact, any charged particles beyond the SM, which are neither too
light~\footnote{Of course, this scenario is also suitable for
lighter particles. However, we can restrict the parameter space (of
the endlessly emerging new models) tighter by other
astrophysical/terrestrial methods. Hence, this scenario may
specialize in new physics particles beyond $\mathrm{TeV}$ energy
scale, especially just above the energy scale the best colliders can
in touch.} nor having a too short lifetime to suffer a Fermi
acceleration, are suitable for our scenario. One immediate example
is $W_1^\pm$ in UEDs, which we do not discuss in this paper because
the extension is really straightforward. $W_1^\pm$ can cascade decay
to $L_1$ or $\nu_1$~\cite{Cheng:2002ab}; however, because of their
analogous masses, the lifetimes of $W_1^\pm$ should be much longer
than $W^\pm$ in the Glashow-Weinberg-Salam theory of weak
interactions. In fact, $W_1^\pm$ is also an intermediate state of
our $g_1$ decay in this paper. Some lineage scenarios of the ADD
model, which allow bulk bosons rather than bulk
fermions~\cite{DeRujula:2000he,ArkaniHamed:1999dc,Dicus:2000hm}, can
also excite Kaluza-Klein $W^\pm$ which suffer our accelerating.
However, a careful calculation of production rates and lifetimes is
absent. There are also a lot of lineage scenarios of the RS model
which allow the bulk SM
fields~\cite{Goldberger:1999wh,Chang:1999nh,Gherghetta:2000qt}.
These models are more reasonable, because the RS model has an
inherent orbifold configuration (to obtain chiral fermions), and the
bulk SM fields can help us to understand some stiff physical
problems like fermion mass hierarchy. Charged sparticles in
supersymmetric (SUSY) models are also good candidates for these kind
of scenarios~\cite{Martin:1997ns}. In order to explain the
non-baryonic dark matter, the lightest sypersymmetric particle (LSP)
is preferred to be electrically neutral; however, it is not supposed
to do so. Even if the LSP is really neutral, they can also be
accelerated by the Fermi mechanism if the lifetimes of the charged
ones decaying to them are not very short. Whereas different from the
case of the LKP $\gamma_1$, if the LSP is gravitino or the lightest
neutralino, they may hardly cause air shower processes because of
their relatively small cross sections, even if they bang into the
earth with tremendous energy. Gluino (which may exist as the form of
gluino-containing hadron, compare to our KK proton) in split
supersymmetry is also a good idea. There has already been one paper
in Ref.~\cite{Anchordoqui:2004bd}, in which the gluinos are
produced by astrophysical $p + p$ collisions; however, (the
astrophysical aspect of) our scenario has a lot of advantages than
theirs, include: (i) One of the protons in their scenario stays at
rest, hence the center-of-mass energy $\sqrt{s} = \sqrt{m_\mathrm{p}
E_\mathrm{p}}$ of $p + p$ collisions is at least
$10^{14-15}\,\mathrm{eV}$, which is no more than $p +
\mathrm{(air)}$ (for UHECRs to collide with the atmosphere hadrons)
center-of-mass energy here in earth; however, because both of the
protons in our scenario are Fermi accelerated, the center-of-mass
energy of our $p + p$ collisions should be at least
$10^{18}\,\mathrm{eV}$ (we have already derived an overall Lorentz
factor $\Gamma \sim 300$), which is impossibility for any other
scenarios to achieve near earth. Despite the fact that we do not
know whether our ideas of (renormalizable) quantum field theory or
new physics nowadays are suitable for such a huge center-of-mass
energy $\sqrt{s}$, we know that something should happen there. (ii)
The maximum energy a gluino-containing hadron can achieve in their
scenario is only $10^{13.6}\,\mathrm{eV}$~\cite{Anchordoqui:2007pn};
however, the maximum energy of our exotic cosmic-ray particles, can
be even larger than the GZK cutoff (see \S\ref{(subsec)_GZK_cutoff}
for a detailed discussion). Hence, because of the fact that the
cosmic-ray spectrum itself has a large negative power law index of
about $- \bar{\alpha} \sim -2.7$ to $-3.0$ (below or above the
``knee'' energy), for gluino-containing hadrons of energy
$E_\mathrm{exo}$ produced by protons with energy $E_\mathrm{p}$,
their content of cosmic-rays with definite $E$ has an additional
inhibitory factor of $(E_\mathrm{exo} /
E_\mathrm{p})^{\bar{\alpha}}$, which has an order of magnitude of
$(10^{13.6}/E_\mathrm{knee})^{2.7} \times
(E_\mathrm{knee}/E_\mathrm{GZK})^{3} \sim 7.7 \times 10^{-19}$. If
most of the UHE protons did not produce gluino-containing hadrons,
the inhibition should be even stronger. Thus even if the
gluino-containing hadrons are recorded by our scientific equipment
(e.g., the cosmic-ray observatories), they are very difficult to be
found out by such a lot of events with similar energy. However,
because our charged exotic particles (KK protons in the context) are
also accelerated by Fermi acceleration, their content in the UHE
region should be as large as $10^{-5}$ to $10^{-2}$, hence not very
hard to be identified. (iii) We do not really need the charged
exotic particles to be longeval enough to suffer the travel from the
source to earth; it is enough that their longevities are long enough
to suffer an astrophysical Fermi acceleration. Hence, we do not have
to worry about some adolescent (comparison with the``baby universe''
era) cosmological bounds, such as the predictions of Big-bang
nucleosynthesis, or cosmic microwave background. Another very
interesting particle candidate for our scenario is the charged
massive particles (CHAMPs)~\cite{DeRujula:1989fe}. It is an
ambitious dark matter constituent as yet (hence it is longeval to
suffer a Fermi acceleration). The association of our astrophysical
scenario drawing in this context and the CHAMPs, is a very interest
issue; however, we leave it to the later publications.

\section*{Acknowledgements}

I would like to thank Zhen-Jun Xiao, Jia-Jun Zhang and Han-Qing
Zheng concerning CTEQ and particle physics, Xiang-Yu Wang for some
suggestions of GRB environments, and my father's technical help with
programming. Margus Ott helped me to improve my Chinglish. Yi-Zhong
Fan read the manuscript and gave some pertinent advice. Zi-Gao Dai
helped me at all times.

\bibliographystyle{style}
\bibliography{../0000_misc,../0709_astro_particles,../0803_ADD_RS_DGP_model}

\end{document}